\documentclass[final, 5p, twocolumn]{elsarticle}

\usepackage[utf8]{inputenc}
\usepackage[T1]{fontenc}
\usepackage[english]{babel}
\usepackage[none]{hyphenat}
\usepackage{etoolbox}

\usepackage{amsmath,amssymb,amsfonts}
\usepackage{amstext, mathrsfs, mathtools, textcomp}
\usepackage{empheq}
\usepackage{array}
\usepackage{bbm}
\usepackage{cuted}
\usepackage{nicefrac}
\usepackage{multirow}

\usepackage[most]{tcolorbox}
\usepackage{booktabs}
\usepackage{makecell}

\usepackage{gensymb}
\usepackage{graphicx}
\usepackage{hyperref}

\usepackage{xcolor}
\usepackage[export]{adjustbox}

\usepackage{listings}
\definecolor{codegreen}{rgb}{0,0.6,0}
\definecolor{codegray}{rgb}{0.5,0.5,0.5}
\definecolor{codepurple}{rgb}{0.58,0,0.82}
\definecolor{backcolour}{rgb}{0.95,0.95,0.92}

\lstdefinestyle{mystyle}{
	backgroundcolor=\color{backcolour},   
	commentstyle=\color{codegreen},
	keywordstyle=\color{blue},
	numberstyle=\tiny\color{codegray},
	stringstyle=\color{codepurple},
	basicstyle=\ttfamily\scriptsize,
	breakatwhitespace=false,         
	breaklines=true,                 
	captionpos=b,                    
	keepspaces=true,                 
	numbers=left,                    
	numbersep=5pt,                  
	showspaces=false,                
	showstringspaces=false,
	showtabs=false,                  
	tabsize=2
}
\DeclareFontFamily{U}{mathb}{\hyphenchar\font45}
\DeclareFontShape{U}{mathb}{m}{n}{<5> <6> <7> <8> <9> <10> gen * mathb
	<10.95> mathb10 <12> <14.4> <17.28> <20.74> <24.88> mathb12}{}
\DeclareSymbolFont{mathb}{U}{mathb}{m}{n}
\DeclareMathSymbol{\rcirclearrow}{0}{mathb}{'367}

\renewcommand{\imath}{\text{i}}

\usepackage[final]{changes}

\graphicspath{{Figures/}}

\definecolor{pansypurple}{rgb}{0.47, 0.09, 0.29}
\definecolor{lincolngreen}{rgb}{0.11, 0.35, 0.02}
\definecolor{internationalkleinblue}{rgb}{0.0, 0.18, 0.65}

\newcommand{\pdiff}[3][\empty]{\ifx\empty#1
	\frac{\partial\,#2}{\partial #3}
	\else
	\frac{\partial^{#1}\,#2}{\partial #3^{#1}}
	\fi}		

\newcommand*{\functiondescription}[4]{%
	\vspace{0.5\baselineskip}
	
	\noindent\hspace{0.035\linewidth}
	\fbox{
		\begin{minipage}{0.85\linewidth}
			\vspace{0.25\baselineskip}
			\begin{center}
				\ifthenelse{\equal{#1}{f}}
				{\function{#2}\\\small(function)}
				{\object{#2}\\\small(class)}
			\end{center}
			\if\relax\detokenize{#4}\relax
			\else
			\relax
			\vspace{-0.5\baselineskip}\noindent
			\ifthenelse{\equal{#1}{f}}
			{\textit{arguments:}}
			{\textit{constructor arguments:}}
			\relax
			\functiondescr{#4}
			\fi
		\end{minipage}
	}
	\vspace{\baselineskip}
	
	\par\noindent\relax
}
\newcommand*{\functiondescr}[1]{%
	\begin{itemize}
		\funcparameter#1\relax
	\end{itemize}
}
\newcommand{\funcparameter}[1]{%
	\ifx\relax#1\empty
	\else
	\vspace{-0.5\baselineskip}
	\item #1
	\relax
	\expandafter\funcparameter
	\fi
}

\newcommand{\TITLE}{Inverse design with flexible design targets via deep learning: Tailoring of electric and magnetic multipole scattering from nano-spheres}

\hypersetup{
	colorlinks,       
	linkcolor=blue,          
	citecolor=blue,         
	filecolor=blue,       
	urlcolor=blue,           
	pdfauthor     = {Estrada-Real et al.},
	pdftitle      = {\TITLE},
	pdfsubject    = {PNFA inverse design},
	pdfcreator    = {pdflatex},
	pdfkeywords   = {}
}

\newcounter{bla}

\journal{Photonics and Nanostructures - Fundamentals and Applications}

\begin{document}
	
	\begin{frontmatter}
		
		
		
		\title{\TITLE}
		
		
		\author[a]{Ana Estrada-Real}
		\author[b]{Abdourahman Khaireh-Walieh}
		\author[a]{Bernhard Urbaszek}
		\author[b]{Peter R. Wiecha\corref{author}}
		\cortext[author] {Corresponding author.\\\textit{E-mail address:} pwiecha@laas.fr}
		\address[a]{INSA-CNRS-UPS, LPCNO, Universit\'e de Toulouse, 31000 Toulouse, France}
		\address[b]{LAAS-CNRS, Universit\'e de Toulouse, 31000 Toulouse, France}
		
		\begin{abstract}
			Deep learning is a promising, ultra-fast approach for inverse design in nano-optics, but despite fast advancement of the field, the computational cost of dataset generation, as well as of the training procedure itself remains a major bottleneck. This is particularly inconvenient because new data need to be generated and a new network needs to be trained for any modification of the problem.
			We propose a technique that allows to train a single neural network on a broad range of design targets without any re-training.
			The key idea of our method is to enrich existing data with random ``regions of interest'' (ROI) labels.
			A model trained on such ROI-decorated data becomes capable to operate on a broad range of physical targets, while it learns to focus its design effort on a user-defined ROI, ignoring the rest of the physical domain.
			We demonstrate the method by training a tandem-network on the design of dielectric core-shell nano-spheres for electric and magnetic dipole and quadrupole scattering over a broad spectral range.
			The network learns to tailor very distinct, flexible design targets like scattering due to specific multipoles in narrow spectral windows. Varying the design problem does not require any re-training. 
			Our approach is very general and can be directly used with existing datasets. It can be straightforwardly applied to other network architectures and problems.
		\end{abstract}
		
		\begin{keyword}
			nano-optics \sep Mie theory \sep core-shell nano-sphere \sep multipole scattering \sep inverse design \sep deep learning
			
		\end{keyword}
		
	\end{frontmatter}
	
	\section{Introduction}
	
	At sub-wavelength scales, various unique optical effects occur as a result of electromagnetic resonances in nanostructures.
	The corresponding field of nano-photonics grew rapidly in the past two decades, enabling unprecedented control of light via specifically designed nanostructures \cite{novotnyPrinciplesNanooptics2006, maierPlasmonicsFundamentalsApplications2010, kuznetsovOpticallyResonantDielectric2016}. 
	With the advancement of nano-photonics, the quest for optimum designs of nano-resonators progressed quickly.
	Unfortunately the inverse design of photonic nanostructures is an ill-posed problem. 
	Generally it cannot be solved analytically, even if an analytical model for the physical response is available. 
	While for simple functionalities and applications, intuitive design choices can be sufficient, optimization methods have proven to deliver significantly better performing nano-structures. 
	Examples for applications of accordingly optimized nano-structures are near-field enhancement, diffraction limited color rendering, nano-scale heat generation or optical directional antennas \cite{feichtnerEvolutionaryOptimizationOptical2012, girardDesigningThermoplasmonicProperties2018, gonzalez-alcaldeOptimizationAlldielectricStructures2018, wiechaMultiresonantSiliconNanoantennas2018, wiechaEvolutionaryMultiobjectiveOptimization2017, wiechaDesignPlasmonicDirectional2019, jiangGlobalOptimizationDielectric2019, elsawyNumericalOptimizationMethods2020, wangAdvancingStatisticalLearning2022}.
	Conventional optimization techniques are based on iterative improvements of a trial solution, thus require large numbers of evaluations (usually computationally demanding simulations). 
	Furthermore, for every single inverse design run the iterative scheme has to be repeated. 
	As a consequence, optimization based inverse design is computationally very expensive.
	
	Recently, deep learning methods have been proposed to inverse design photonic nanostructures significantly faster \cite{malkielPlasmonicNanostructureDesign2018, hegdeDeepLearningNew2020, wiechaDeepLearningNanophotonics2021, jiangDeepNeuralNetworks2021}.
	However, while the inverse design usually indeed is very fast, deep learning techniques come with a considerable computational overhead. The neural networks require large amounts of training data, and the training step itself is computationally heavy.

	\begin{figure*}[t]
		\centering
		\includegraphics[width=0.85\linewidth]{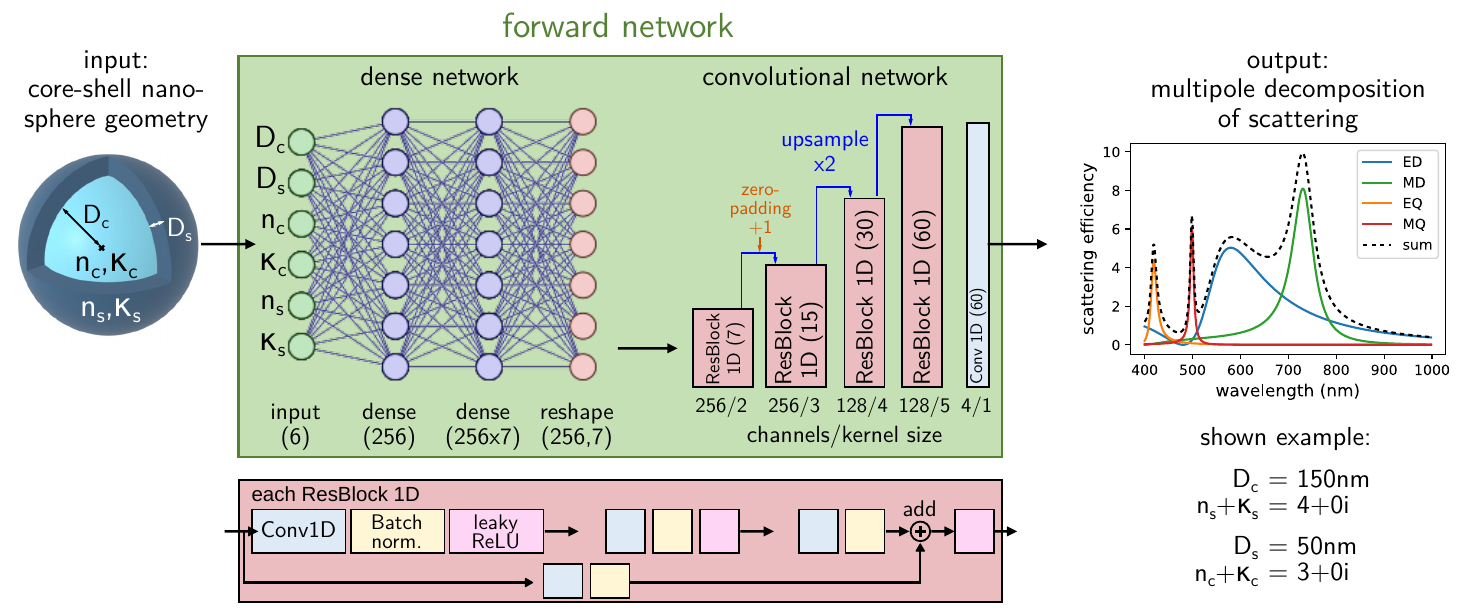}
		\caption{
			Our forward neural network predicts an approximation for the scattering efficiency of dielectric core-shell nano-spheres, decomposed into the first multipole orders: electric dipole (ED, blue), magnetic dipole (MD, green), electric quadrupole (EQ, orange) and magnetic quadrupole (MQ, red).
			The network takes as input six values: The core and shell thickness ($D_c$ and $D_s$) as well as their complex refractive indices (real and imaginary parts $n$, respectively $\kappa$).
			The 1D convolutional part that follows a dense network, is composed of residual blocks (ResBlock, details shown in the red box at the bottom), the output is a convolutional layer with four channels and kernel size 1, that returns the spectra of the multipole contributions to the scattering.
			The shown example geometry is given under the spectra on the right.
		}
		\label{fig:ann_forward_scheme}
	\end{figure*}

	In addition to the expensive data generation and training, the ill-posed character of typical inverse design problems requires adequate, relatively complex network architectures \cite{wiechaDeepLearningNanophotonics2021, dengNeuraladjointMethodInverse2021, renInverseDeepLearning2022, chenArtificialIntelligenceMetaoptics2022}.
	Inverse design-compatible network types are for instance conditional generative adversarial networks (cGAN), conditional adversarial autoencoders (cAAE), conditional variational autoencoders (cVAE) or invertible networks \cite{liuGenerativeModelInverse2018, soDesigningNanophotonicStructures2019, dinsdaleDeepLearningEnabled2021, kudyshevMachineLearningAssisted2021, ardizzoneAnalyzingInverseProblems2018, blanchard-dionneSuccessiveTrainingGenerative2021}.
	Hyper-parameter tuning for such sophisticated models is a non-trivial task by itself and obtaining good training convergence can be tedious.
	A relatively simple, yet very efficient inverse design network architecture is the so-called \textit{tandem network}, which is a variant of an autoencoder \cite{kingmaIntroductionVariationalAutoencoders2019} consisting of a pre-trained, physics predicting decoder (the ``forward'' network), which stabilizes training of the actual inverse network (the encoder) \cite{malkielPlasmonicNanostructureDesign2018,  liuTrainingDeepNeural2018}.
	
	Once trained, an inverse design neural network is fabulously fast, delivering its inverse design predictions in fractions of seconds, often even in fractions of milliseconds.
	The main drawback of deep neural networks for inverse design is hence the expensive training step (including data generation). 
	Unfortunately, the geometric model and the design problem are implicitly defined by the training data and the network layout. 
	Consequently, a new network needs to be trained on a new data-set, as soon as either the geometry or the problem changes. 
	While the geometry can be easily generalized using free-form parametrizations e.g. via pixelization or voxelization \cite{wenRobustFreeformMetasurface2020, wiechaDeepLearningMeets2020}, the target properties (e.g. reflectivity or scattering) or for instance their wavelength range are generally hard-coded \cite{dinsdaleDeepLearningEnabled2021, liuTrainingDeepNeural2018, soSimultaneousInverseDesign2019}.
	Every change of the objective hence requires re-training of a new network.

	\begin{figure*}[t]
		\centering
		\includegraphics[width=\linewidth]{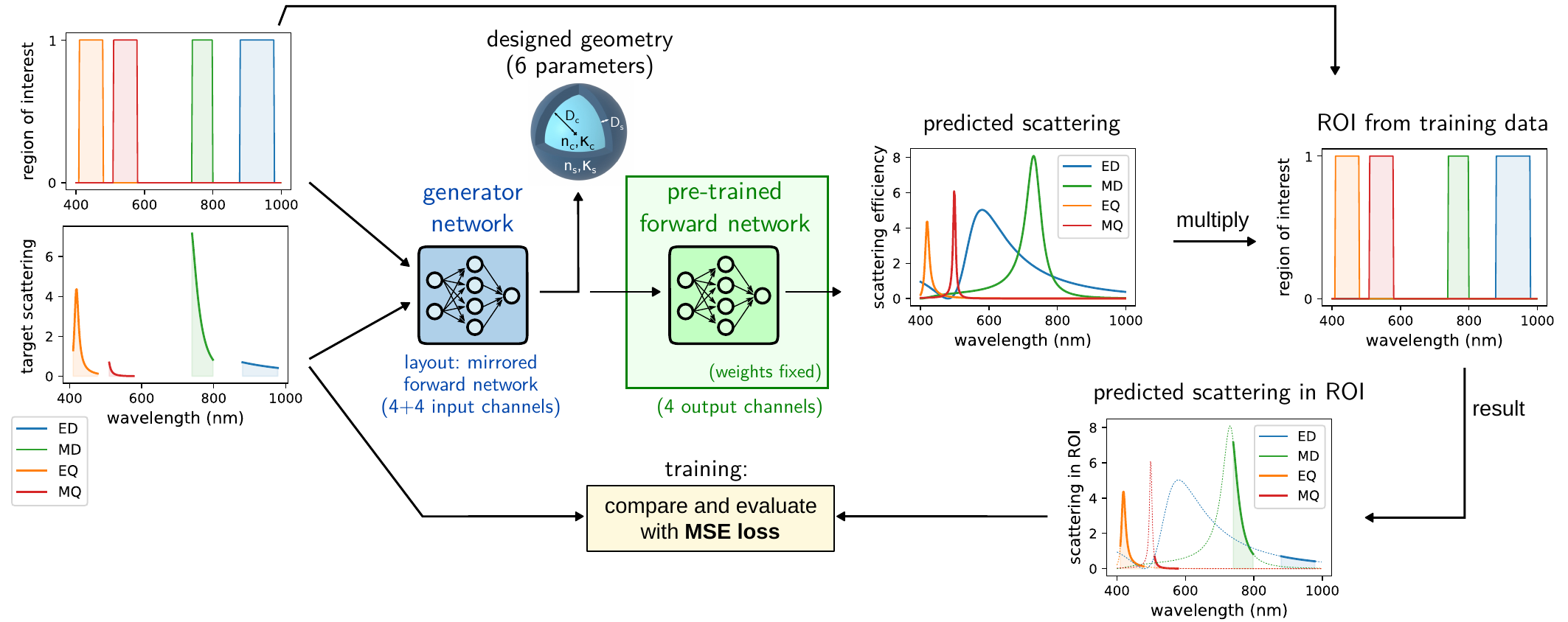}
		\caption{
			Inverse network scheme. 
			The actual generator network takes as input the spectral region of interest (ROI) as well the target scattering in the ROI for each multipole (4+4 inputs). The goal is to return the geometry of a core-shell nano-sphere that yields the target spectrum.
			During training, this predicted design is fed back into the pre-trained forward network (see Fig.~\ref{fig:ann_forward_scheme}), whose parameters are fixed and which thus is not subject to further training.
			The predicted physical response of the inverse designed nano-sphere is subsequently multiplied with the target ROI of the training sample. 
			The scattering spectra inside the ROIs are finally compared via a mean square error metric, which is minimized during training.
		}
		\label{fig:ann_tandem_scheme}
	\end{figure*}

	To resolve this limitation of current models, here we propose a variation of the tandem network, that allows to flexibly adapt the inverse design objective at the inference stage, hence without requiring expensive re-training.
	Our modification consists in including an additional ``region of interest'' input to the design network, which allows to specifically define the inverse design objective, for example by setting a narrow wavelength window or by choosing one of several possible observables.
	Our approach requires only a modification of the network architecture and its training sequence, it can thus be used on existing data without the need of additional simulations.
	We illustrate the concept by the example of inverse design of electric and magnetic dipole and quadrupole optical resonances in dielectric core-shell nano-spheres of sizes up to a few hundreds of nanometers.
	Such core-shell particles occur in atmospheric aerosols and have been suggested for applications based on directional scattering or for cloaking \cite{liuScatteringCoreshellNanowires2013, h.jonesAtmosphericallyRelevantCore2015, liuToroidalDipoleinducedTransparency2015, fengIdealMagneticDipole2017}. 
	They are also a popular benchmark platform for inverse design, since light scattering can be analytically calculated via Mie theory, while the multiple spherical layers offer a broad design flexibility \cite{soSimultaneousInverseDesign2019,  peurifoyNanophotonicParticleSimulation2018, huRobustInversedesignScattering2019, sheverdinPhotonicInverseDesign2020}.
	To benchmark our approach, we provide a statistical evaluation of the network's accuracy and demonstrate the remarkable flexibility on a wide range of design targets.

	\section{Design problem, data generation, deep learning models}
	
	\paragraph{Physical problem}
	
	We choose as physical observable for a benchmark of our approach the multipole-decomposed scattering of dielectric core-shell nano-spheres.
	The interplay of electric and magnetic modes can lead to interesting effects such as uni-directional forward scattering due to the Kerker-effect or suppression of scattering while maximizing the energy density inside a nano-sphere as a result of anapole states \cite{kerkerElectromagneticScatteringMagnetic1983, fuDirectionalVisibleLight2013, staudeTailoringDirectionalScattering2013, liuScatteringCoreshellNanowires2013, miroshnichenkoNonradiatingAnapoleModes2015, wiechaStronglyDirectionalScattering2017}.
	Practically, together with homogeneous spheres, core-shell nano-spheres are the most relevant spherical nano-particles. 
	Precise chemical synthesis of nano-spheres with more than one coating layer is increasingly difficult, most practical applications are therefore limited to homogeneous structures or core-shell geometries \cite{marcoBroadbandForwardLight2021}.
	
	Using Mie theory, we calculate the electric and magnetic dipole and quadrupole contributions to the scattering efficiency, which is defined as ratio of scattering cross section and geometric cross section. 
	As an example, on the right of figure~\ref{fig:ann_forward_scheme} we show the spectra of a non-absorbing core-shell nano-sphere with core refractive index $n_c=4$ (materials with similar index in the visible would be crystalline silicon or GaAs) and shell refractive index of $n_s=3$ (e.g. HgS).
	For simplicity we use constant refractive indices for core and shell, but dispersive materials may be easily treated, e.g. using indexed lookup tables \cite{soSimultaneousInverseDesign2019}.
	For the numerical calculation we use the python package ``PyMieScatt'' \cite{sumlinRetrievingAerosolComplex2018}.
	
	The inverse problem that we seek to solve is to find the core and shell thickness as well as their refractive indices, in order to yield specific scattering efficiencies due to one or several selected multipoles and within user-defined, variable spectral windows.

	\paragraph{Iterative optimization vs. one-shot inverse network}
	
	Conventional inverse design uses an iterative algorithm to optimize a nanostructure for a specific task \cite{elsawyNumericalOptimizationMethods2020}. This approach has two key advantages: First the design target can be entirely freely defined. And second, the use of an optimization algorithm means that the procedure converges to a close-to-optimum solution for the problem. The main drawback of the technique is speed: The iterative search usually requires hundreds or thousands of structure evaluations. Furthermore this expensive search has to be repeated from scratch for every design target.
	
	Inverse design neural networks on the other hand need to be trained on a large dataset, requiring an initial, large computational invest. From then on, they operate in a ``one-shot'' manner: a single, very rapid evaluation yields a design suggestion almost instantaneously \cite{wiechaDeepLearningNanophotonics2021}. A major drawback is here, that one cannot be sure if the design is optimum.
	
	A hybrid solution can be the use of a forward neural network as a fast surrogate model for slow simulations, to accelerate iterative optimization \cite{hegdeDeepLearningNew2020}. A risk of such approach is that the search for extrema in the design space can converge to singularities of the neural network model.

	\paragraph{Forward network}
	The first step in creating the inverse design model is the training of a forward network. This neural network takes as input the nano-sphere geometry and predicts an approximation for the physical response, in our case the spectra of scattering of the nano-sphere, decomposed in electic dipole (ED), magnetic dipole (MD), electric quadrupole (EQ) and magnetic quadrupole (MQ) contributions. 
	A detailed illustration of the network architecture is shown in figure~\ref{fig:ann_forward_scheme}.
	It consists of a first, dense network part, followed by an upsampling 1D convolutional neural network (CNN). The CNN is built from four residual blocks (resBlocks) \cite{heDeepResidualLearning2015, szegedyInceptionv4InceptionResNetImpact2016}, each composed of 3 convolutions and a skip connection. 
	Each resBlock is followed by an upsampling layer to expand the data size of 7 values at the first convolution to the final size of 60 wavelengths for each of the four spectra at the network's output. 
	Zero padding is used after the first upsampling for dimensionality matching.
	Through the network, Leaky ReLU is used as activation \cite{maasRectifierNonlinearitiesImprove2013}, except for a linear activation at the output.
	
	
	The training data consists of a large set of random core-shell geometries and their scattering spectra, calculated by Mie-theory.
	We separate the contributions to scattering of the first four multipole modes. 
	We generate a dataset of 20,000 random nano-spheres and their multipole scattering spectra. 
	The core and shell thicknesses are randomly chosen between $50\,$nm and $200\,$nm. 
	The real part of the refractive indices $n$ lie between $2.0$ and $4.0$, the limits of the extinction coefficients $\kappa$ are set as $0.0$ and $0.25$.
	The layer thicknesses and refractive indices are separately normalized using a standard scaler \cite{goodfellowDeepLearning2016}. 
	The spectra are already normalized to the nano-sphere's geometric cross section, thus no further normalization is done.
	Of this dataset 18,000 samples are used for training and 2,000 for testing. 
	The forward network is then trained on the simultaneous prediction of all four multipole spectra. 
	Training is done using a mean square error (MSE) loss with the ADAM optimizer \cite{kingmaAdamMethodStochastic2014}.
	It runs over 600 epochs, and the learning rate is successively decayed from a starting value of $10^{-3}$ to $10^{-6}$. 
	
	In the tandem architecture that we will use in the following, the forward network is used as surrogate physics predictor during training of the inverse network. Therefore, an accurate forward model is particularly crucial \cite{wiechaDeepLearningNanophotonics2021}.
	The median relative prediction error on 2,000 test samples in the case of our core-shell multipole scattering forward network, is as low as $0.53$\%, which is an excellent starting point for training of the inverse model.

	\paragraph{Generator network}
	
	In the subsequent step, we use the working forward network to train a generator on the task of inverse designing nano-spheres with a user-defined optical response.
	Since potentially several solutions exist for every design target, the generator training loss is calculated by comparing the target spectra with the forward network's predicted spectra.
	In this way, a comparison of the design parameters is not required during training, and possible non-unique solutions do not deteriorate the training convergence \cite{liuTrainingDeepNeural2018}. 
	
	So far the described approach corresponds to the conventional tandem network \cite{wiechaDeepLearningNanophotonics2021}. 
	A main limitation of the latter is, that the design target is hard-coded in the network layout.
	In our case, tailoring a specific optical response requires the definition of the target over the full spectral region, covered by the network \cite{soSimultaneousInverseDesign2019}. 
	It would be furthermore necessary to define the target scattering for every single multipole, even if for example only the magnetic dipole response is of interest.
	
	Therefore the principal contribution of our work consists in a modification that allows to flexibly restrict the design target during network evaluation via additional ``region of interest'' (ROI) input channels. This is illustrated in figure~\ref{fig:ann_tandem_scheme}.
	These ROI channels tell the network for every physical property, the range inside which we aim to tailor the optical response. 
	In our example the ROIs defines for every multipole a spectral window that shall be considered for inverse design, while the spectral regions where the ROI is set to zero are to be ignored for the design. 
	Please note that in computer vision applications, ROIs are often defined as bounding box coordinates \cite{girshickFastRCNN2015, minaeeImageSegmentationUsing2020}. Here however, similar to image segmentation applications, the ROIs have the same dimensionality as the spectral input, offering a larger flexibility at the cost of a larger dataset memory requirement.
	In this work we consider the first four multipole orders, therefore our generator has $4+4$ inputs.
	Apart from the additional inputs, the generator is an exact mirror version of the forward network, where the \textit{reshape} layer is replaced by \textit{flatten}, and the \textit{upsampling} layers by \textit{maxpooling} operations.
	
	Both networks are of moderate size. The forward network has around 2.5 million trainable parameters, the generator around 2 million. The smaller amount for the generator stems from the max pooling layers, after which the subsequent convolution requires only half the parameters compared to the equivalent upsampling operation.

	In addition to the ROI input channels for the generator, we also need to generate appropriate training data by pre-processing the Mie-calculated spectra. 
	We do so by generating random ROIs for multiplication with each multipole spectrum. 
	These random ROIs have a spectral width between $50\,$nm and $300\,$nm, and are positioned at a random wavelength. 
	Furthermore, with a probability of $20$\,\%, we set the ROI over the entire wavelength range to $1$, and with another $20$\% chance, we set it entirely to zero. 
	An example of a Mie-spectrum combined with a random set of ROIs is shown on the left of figure~\ref{fig:ann_tandem_scheme}.
	Since the ROIs are randomly generated, we can use every sample of our ``physical'' dataset several times, each time with a different ROI. Here, we generate 100,000 ROI-scattering samples from our 20,000 core-shell nano-spheres.
	
	The last modification, compared to the vanilla tandem network, concerns the training loss calculation. 
	The forward network predicts the full response of the design, but we are interested only in the scattering inside the ROIs. 
	We therefore use the ROI input channel in order to multiply the forward network's prediction with the target ROI. 
	The MSE training loss is finally calculated between target scattering and predicted scattering inside the ROI. 
	The training procedure is illustrated in figure~\ref{fig:ann_tandem_scheme}.

	\begin{figure}[t]
		\centering
		\includegraphics[width=\linewidth]{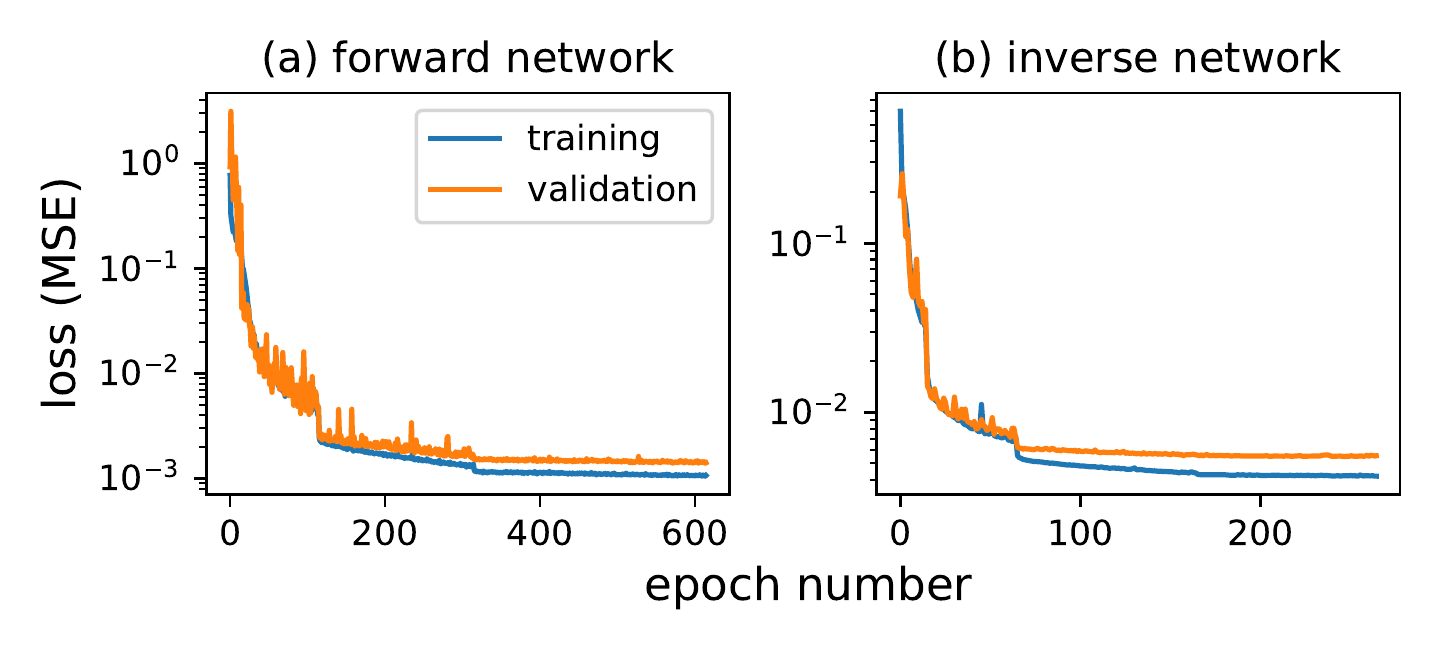}
		\caption{
			Training convergence of (a) the forward network, and (b) the tandem inverse network trained on the ROI data.
		}
		\label{fig:training_progress}
	\end{figure}
	
	We train the networks on an NVIDIA RTX 2060 SUPER GPU with 6 GByte RAM. Training of the forward network takes around $5$\,s per epoch (roughly 1 hour for full training), and $32\,$s per epoch for the inverse design tandem network (total training of around 3 hours).
	Inference on the same GPU takes (per sample) around $90$\textmu s for the forward network and $110$\textmu s for the generator network.
	The (CPU-based) Mie calculations for one spectrum require around 60\,ms. Comparing this to the GPU-evaluated forward neural network (around 0.1ms per spectrum) corresponds to a forward-acceleration by a factor of around 600.
	Please note that the Mie code is not parallelized and doesn't run on GPU as the neural network does, therefore the comparison is not exactly objective.

	Figure~\ref{fig:training_progress} shows the evolution of the loss during training of the forward model (\ref{fig:training_progress}a), respectively the inverse network (\ref{fig:training_progress}b), demonstrating that the training converged and no overfitting occurred.

	\begin{figure}[t]
		\centering
		\includegraphics[width=0.8\linewidth]{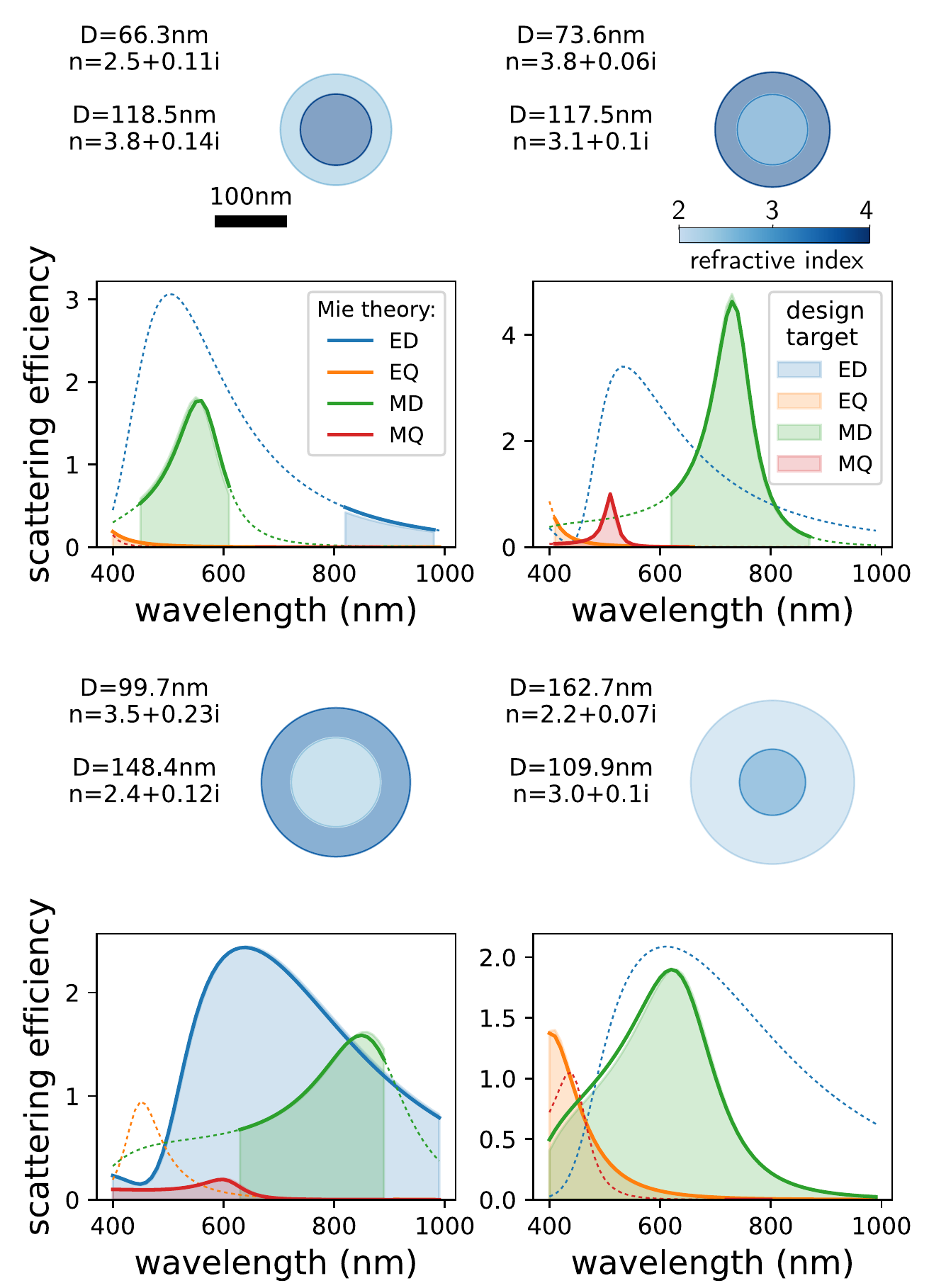}
		\caption{
			Selected inverse design examples using samples from the validation set as targets. 
			The shaded regions indicate the inverse design goal.
		}
		\label{fig:validation_examples}
	\end{figure}

	\section{Results}

	In inverse design via iterative optimization, a solution to the design problem is well defined, namely as the minimum deviation between design target and optimized geometry property. This definition still holds if it is physically not possible to exactly achieve the actual target, then the design solution is the closest possible match \cite{elsawyNumericalOptimizationMethods2020}.
	
	Data-driven techniques like neural networks on the other hand, are trained on simulated samples of which the calculated optical response is used as the design target for the training set. Neural networks are hence generally trained only on ``exact'' solutions for the design targets. 
	As a result it is in general not foreseeable how an inverse network performs on arbitrary targets where no exactly matching solution exists.
	Without performing a conventional simulation of the final result, there is no way to determine if a solution is in fact optimum, if it is even close to the desired result, or possibly entirely wrong. 
	It is possible to use a forward network to assess the design performance, however then the reliability problem is merely passed onto another neural network model, which may as well suffer from unpredictable outliers, from weak generalization outside the training parameter space, or even from singularities where the model totally collapses \cite{huangAdversarialAttacksNeural2017, wiechaDeepLearningMeets2020, liuAdversarialAttacksDefenses2021}.
	
	Our approach to verify the trustworthiness of the model is to perform very extensive and systematic tests of the network's fidelity.
	We first perform a statistical benchmark, and subsequently systematic tests of different manually defined design targets.

	\subsection{Test-samples and statistics of network performance}
	
	After training, we test the network first on a few random samples from the test-set. 
	The goal is to find a core-shell design that reproduces the optical properties of a randomly generated reference nano-sphere. 
	There thus exists a solution that perfectly reproduces the requested optical responses.
	We test the design fidelity of our network on a data-set of 10,000 random samples, generated in the same way as the training data. 
	
	In figure~\ref{fig:validation_examples} we show four random examples of the test set. 
	Solid lines correspond to the Mie-calculated scattering of the suggested inverse design in the target region of interest. Thin dashed lines show the spectra outside the design target spectral regions.
	Areas shaded with the respective colors correspond to the design targets for electric dipole (ED, blue), electric quadrupole (EQ, orange), magnetic dipole (MD, green), and magnetic quadrupole (MQ, red). 
	In this qualitative comparison we find an excellent agreement between the design targets and the inverse design Mie spectra.

\begin{figure}[t]
		\centering
		\includegraphics[width=0.9\linewidth]{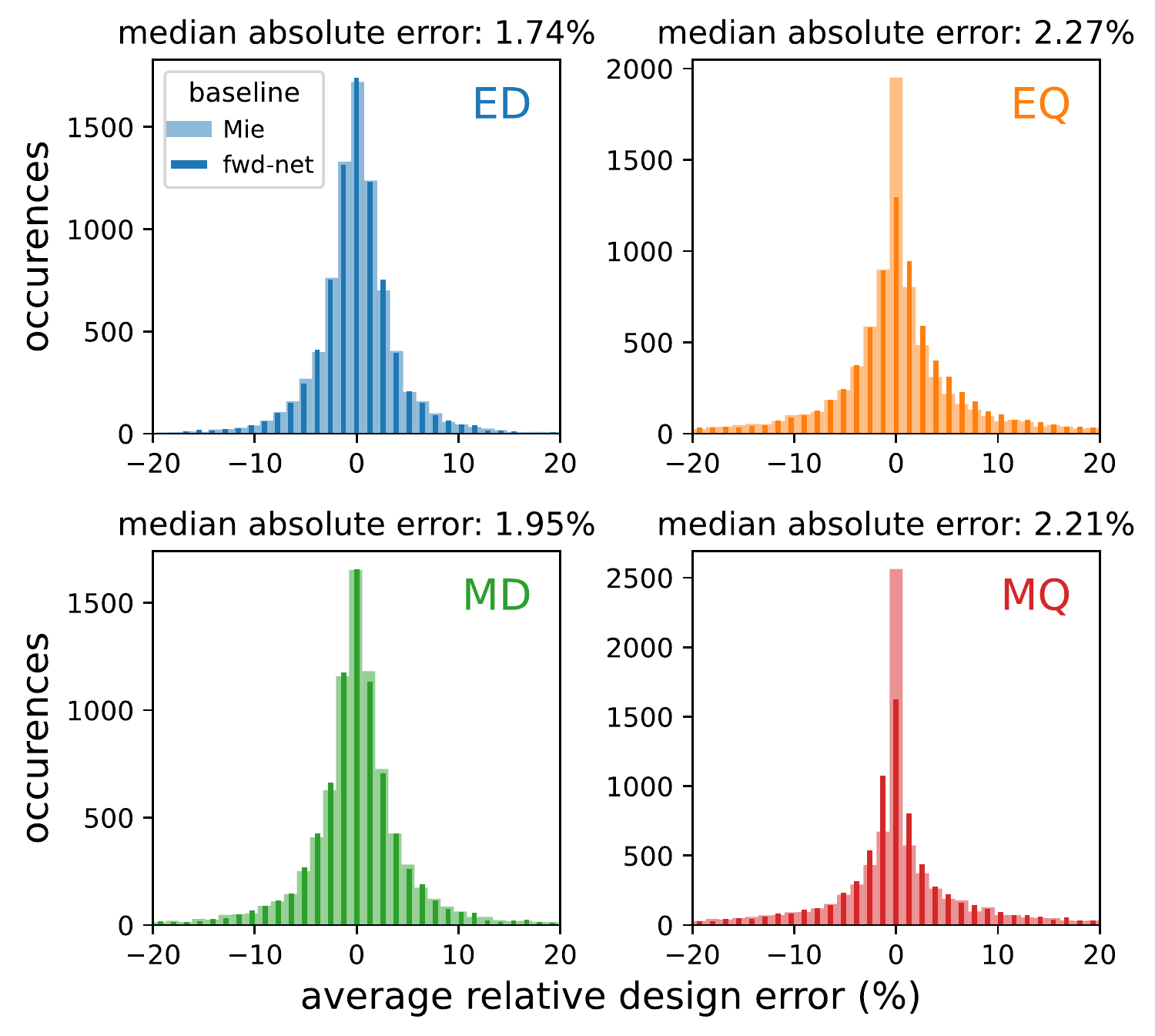}
		\caption{
			Histograms of the inverse network design accuracy for each multipole. 
			The benchmark is performed on the validation set of 10,000 samples. 
			The design error is given in percent. 
			Thick, light shaded bars correspond to the error relative to Mie theory. The thin, darker bars give the accuracy relative to the forward network's prediction.
		}
		\label{fig:statistics_inverse_net}
	\end{figure}

	To assess the design performance in a more quantitative way, Figure~\ref{fig:statistics_inverse_net} shows histograms of the average design accuracy separately for each multipole. 
	The error corresponds to the mean relative deviation between the target spectra and the Mie scattering spectra of the network-designed nano-spheres.
	The thin, solid lines correspond to the predictions of the forward network, while the larger bars of lighter color-shades correspond to the statistics using Mie theory as reference. 
	We find that the forward network's histogram is almost identical to the Mie-calculated errors, indicating the very good performance of the forward net. 
	As mentioned before, this is a crucial condition for robust training of a tandem network \cite{liuTrainingDeepNeural2018, dinsdaleDeepLearningEnabled2021}.
	While the median forward network error was in the order of $0.5$\%, the median of the design errors' absolute values is in the range of 2\% (see labels in figure~\ref{fig:statistics_inverse_net}).

	\subsection{Inverse design of hand-drawn spectra}
	
	After verification of the network's accuracy on pre-calculated nano-spheres in the previous section, we are interested in the capabilities of the inverse network to design ``manually specified'' optical responses, hence lineshapes that are not necessarily physically realistic. 
	The particular challenge is hence, that there no longer exist exact solutions to the given design target. 
	We rather seek the geometry that matches the requested optical response as closely as possible.

	\subsubsection*{Inverse design of relative resonance positions}
	
	\begin{figure*}[p]
		\centering
		\includegraphics[width=\linewidth]{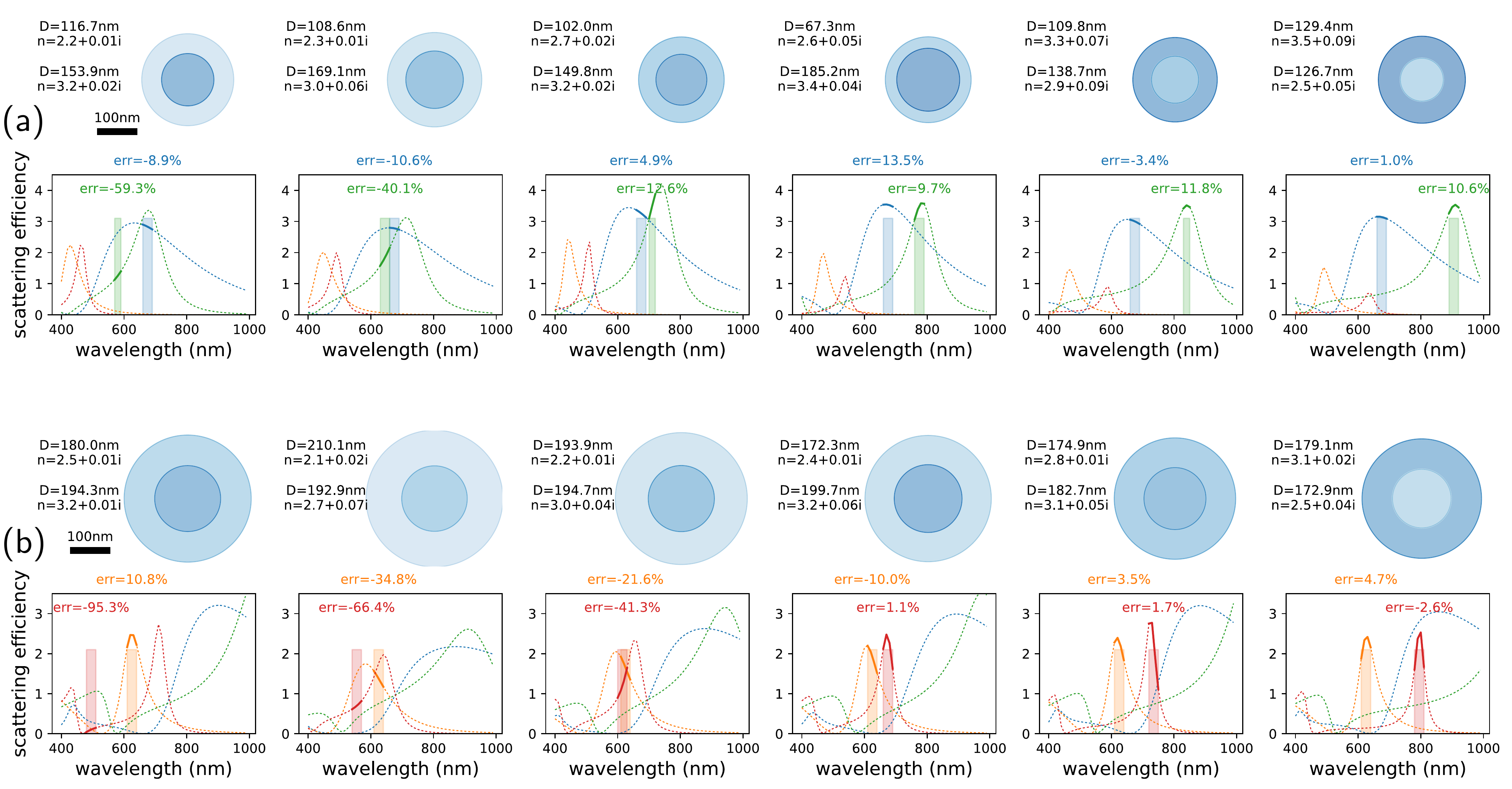}
		\caption{
			Inverse design examples: Design the relative resonance positions of (a) electric and magnetic dipole scattering and (b) electric and magnetic quadrupole scattering.
		}
		\label{fig:examples_relative_resonance_pos}
	\end{figure*}
	
	\begin{figure*}[p]
		\centering
		\includegraphics[width=\linewidth]{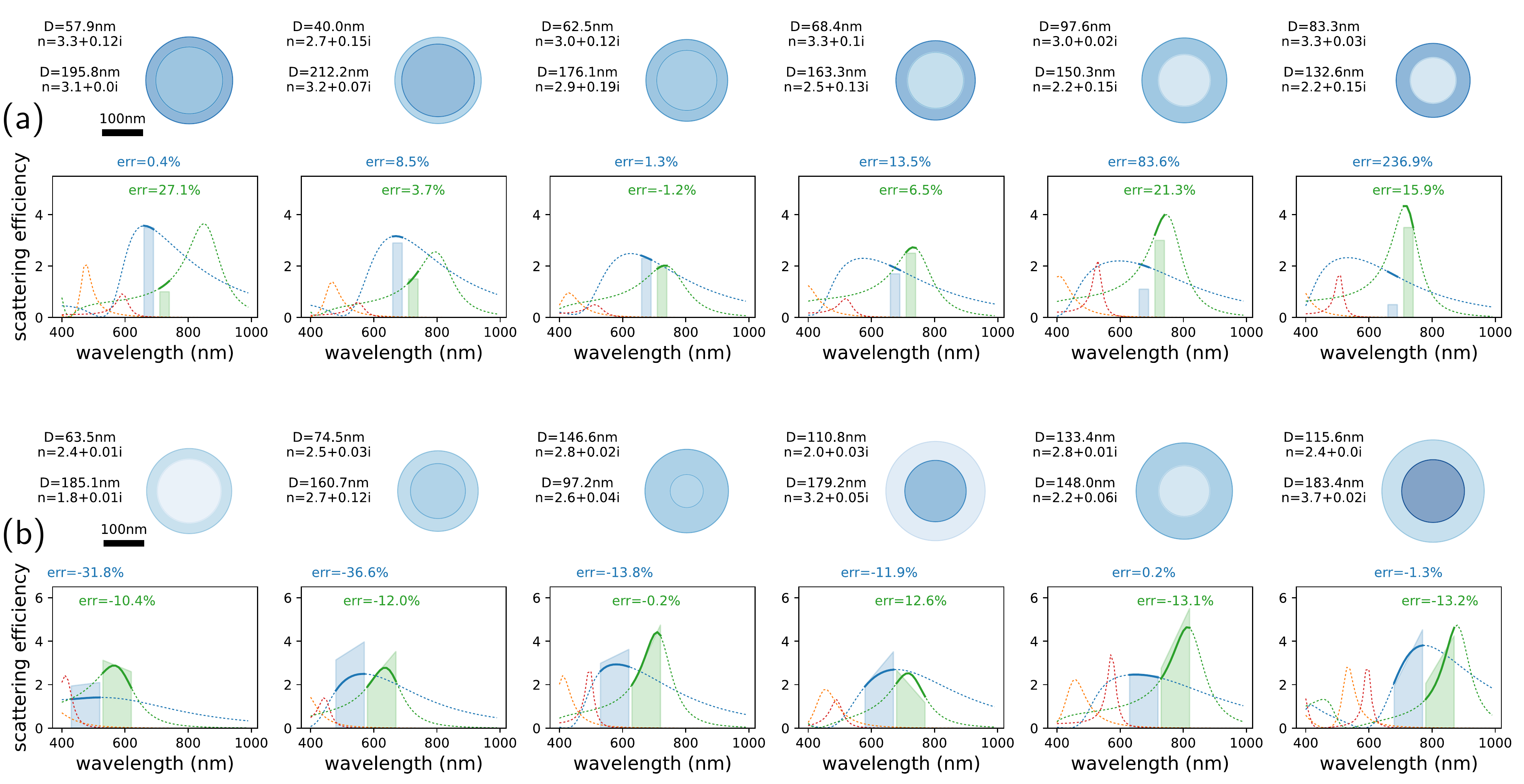}
		\caption{
			Inverse design examples: Tailoring of (a) the relative scattering intensity ratio between electric and magnetic dipole scattering or (b) their spectral slopes.
		}
		\label{fig:examples_relative_intensities}
	\end{figure*}
	
	\begin{figure*}[t]
		\centering
		\includegraphics[width=\linewidth]{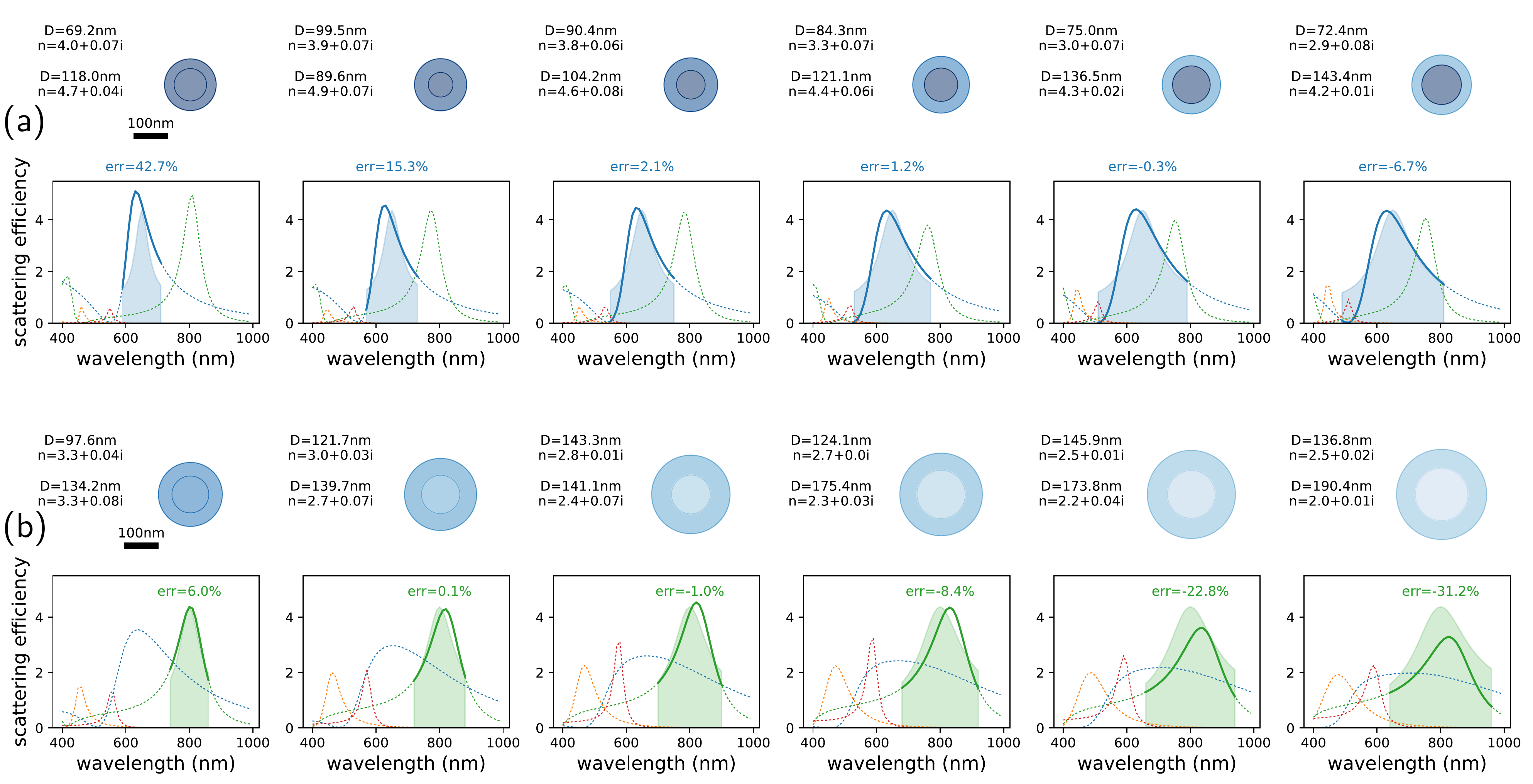}
		\caption{
			Inverse design examples: Design of (a) the electric and (b) the magnetic dipole scattering resonance width.
		}
		\label{fig:examples_resonancewidth_dipole}
	\end{figure*}
	
	\begin{figure*}[t]
		\centering
		\includegraphics[width=\linewidth]{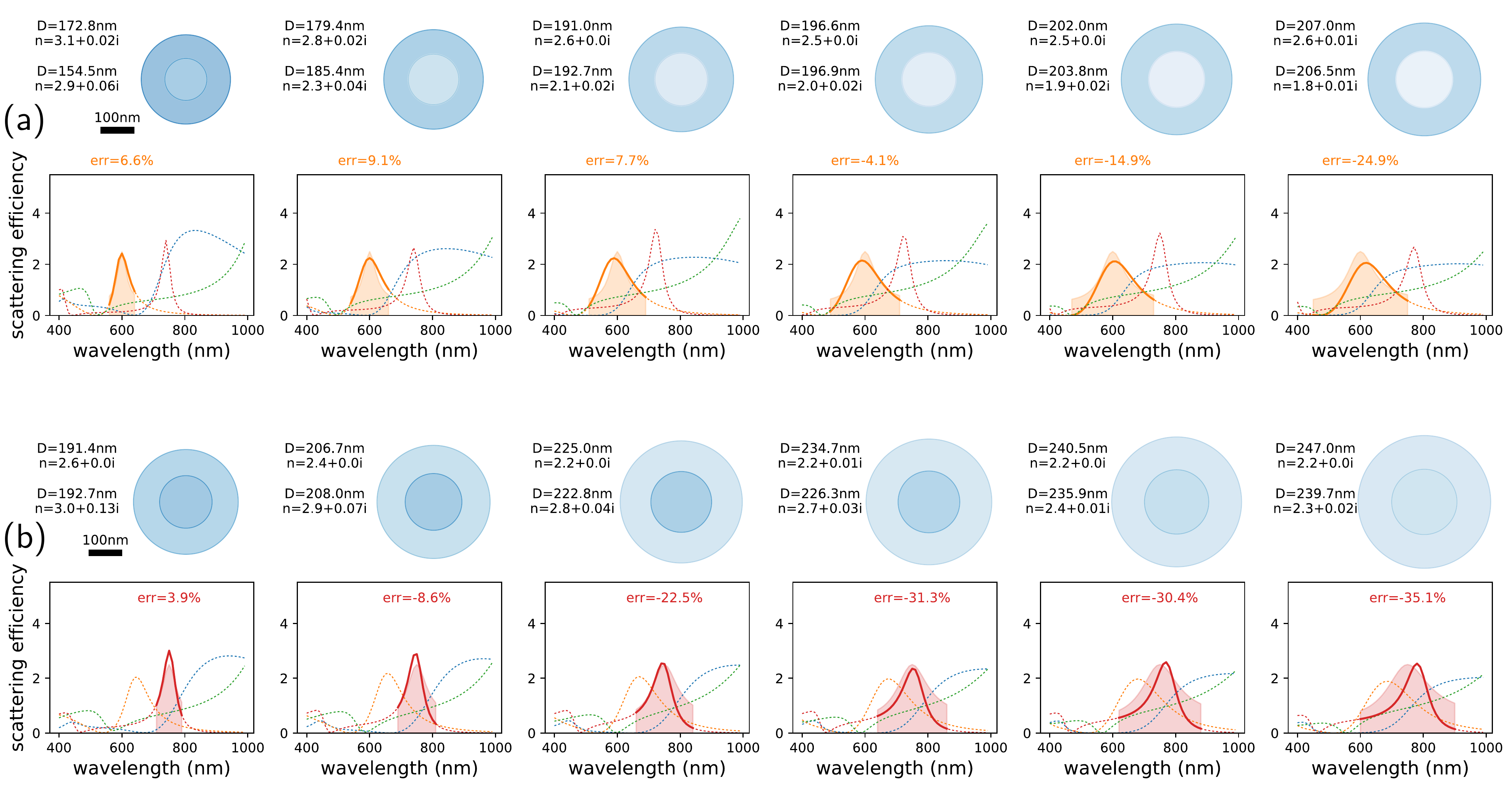}
		\caption{
			Inverse design examples: Design of (a) the electric and (b) the magnetic quadrupole scattering resonance width.
		}
		\label{fig:examples_resonancewidth_quadrupole}
	\end{figure*}

	As a first systematic test, we try to independently control the wavelength for scattering due to the electric and magnetic dipole resonances. 
	To this end, we set the target scattering efficiency to a value of $3$. We aim to design a nano-sphere which has this target scattering efficiency for the electric dipole in a wavelength range between $700$\,nm and $750$\,nm, while we tune the target wavelength of the magnetic dipole of same magnitude from $550$\,nm to $900$\,nm.
	The resulting geometries as well as the corresponding Mie-calculated spectra are shown in figure~\ref{fig:examples_relative_resonance_pos}a.
	For the transition from shorter to longer wavelengths of the MD relative to the ED, the network suggests core-shell nano-spheres transiting from high-index core and lower index shell to high-index shell with a lower index core.
	We note that the solution with lower index shell and higher index core for spectrally overlapping ED and MD are in agreement with recently proposed designs \cite{marcoBroadbandForwardLight2021}.
	We observe that the network doesn't manage to find geometries which have a magnetic dipole resonance at shorter wavelengths than the electric dipole resonance, which is the expected physical behavior for dielectric nanostructures \cite{caoTuningColorSilicon2010, kuznetsovOpticallyResonantDielectric2016}. 
	However, as soon as the MD is set at wavelengths longer than the ED, the inverse network successfully finds suitable approximate solutions.
	
	In figure~\ref{fig:examples_relative_resonance_pos}b we repeat the same test with the electric and magnetic quadrupoles (EQ and MQ). 
	We fix the EQ target scattering efficiency at $2$ in a wavelength window between $600$\,nm and $650$\,nm, then we move the target MQ of same strength between $500$\,nm and $800$\,nm.
	Again, the MQ occurs naturally at longer wavelengths than the EQ, and accordingly the network struggles to find solutions with MQ at shorter wavelengths than the EQ. However, as soon as the MQ is on the red spectral side relative to the EQ, the inverse network finds suitable geometries.

	\subsubsection*{Inverse design of relative multipole strength and spectral slopes}

	In a second systematic test, we now try to vary the relative strength of the electric and magnetic dipole, while keeping the target wavelengths fixed. 
	The inverse designed core-shell sphere geometries and their according Mie calculated scattering spectra are shown in figure~\ref{fig:examples_relative_intensities}a.
	We find that the neural network manages to design nano-spheres with correct scattering efficiency ratios for almost all targets. 
	Physical limitations become visible only for very small electric dipole intensities. 
	By looking at the Mie-spectra outside the design region of interest, we observe that the network does not always use the resonance peaks as target scattering efficiencies, but it often uses the shoulder of the Lorentzian line-shape, which allows an additional degree of freedom in the design flexibility. 
	In order to force the network to design a resonance with maximum at the target wavelength, we will use below a Lorentzian as scattering target.
	
	Since we found the network to use the spectral shoulder to tailor specific scattering intensities, we want to test if it manages to find structures that create user-defined slopes in the scattering spectra.
	Designing the slope of spectral features of the scattering is particularly interesting for sensing applications \cite{meschNonlinearPlasmonicSensing2016, barredaRecentAdvancesHigh2019}.
	We show this in figure~\ref{fig:examples_relative_intensities}b on random examples, where the slope of electric and magnetic dipole scattering is independently target of the inverse design in various spectral regions from the blue to the near infrared. 
	To remain in a physically feasible regime, we aim for MD at longer wavelengths relative to the ED.
	The network manages to inverse design almost all targets with a good fidelity.

	\subsubsection*{Inverse design of resonance lineshape}
	
	In several of the above examples as design target we fixed the scattering efficiency in a narrow spectral window. 
	We found that, in this case, the design network not necessarily defines structures with the peak of the resonances at the selected wavelength.
	In order to demonstrate that it is possible to inverse design also the actual resonance position, we show in figure~\ref{fig:examples_resonancewidth_dipole} Lorentzian line shapes as design targets \cite{soSimultaneousInverseDesign2019}. 
	Figure~\ref{fig:examples_resonancewidth_dipole}a shows inverse design of an electric dipole resonance with increasing linewidth.
	Interestingly, we find that for the most narrow ED line-shapes, the network finds solutions that resemble rather a Fano-profile than a Lorentzian line-shape. This indicates the presence of an Anapole mode at the minimum of the ED scattering efficiency.
	Furthermore, we find here that the network succesfully generalizes outside of the training parameter range, which contained only refractive index real parts between $n=2$ and $n=4$. But here the network uses refractive indices for the core material of up to $n\approx 5$, leading effectively to the desired, narrow lineshapes for the ED scattering.
	Figure~\ref{fig:examples_resonancewidth_dipole}b shows the same increasing width targets but now using the magnetic dipole scattering.
	Again the network manages successfully to tailor the resonance width, by lowering the refractive index while increasing the size of the nano-sphere. 
	Note that we observe that in contrast to the ED width design, where the core index was kept larger than the shell index, increasing the MD resonance width requires a lower core refractive index relative to the shell material.
	
	In figure~\ref{fig:examples_resonancewidth_quadrupole} we repeat the design of the resonance width using as target the electric (\ref{fig:examples_resonancewidth_quadrupole}a) and magnetic quadrupoles (\ref{fig:examples_resonancewidth_quadrupole}b).
	Because of the in general weaker quadrupolar scattering, we reduce the target peak scattering efficiency from $4$ for the dipoles to $2.5$ for the quadrupoles.
	The network manages also for the EQ and MQ targets to successfully tailor the resonance width. However, in particular for the broadest design targets it struggles to find appropriate core-shell spheres. 
	We interpret this as a physical limitation of our geometric model, since the quadrupolar resonances (in particular the MQ) generally have narrower line widths compared to the dipole resonances.

	\section{Conclusions}
	
	In conclusion, we presented a method to render deep learning based inverse design more flexible, by including additional channels in the generator, which serve for defining the inverse target.
	Our technique is based on a dynamical restriction of the physical response to a flexible region of interest.
	It can be used directly with existing data-sets and can be implemented with any inverse design network architecture.
	It could be implemented for example with conditional variational autoencoders (cVAE) that allow to distinguish multiple non-unique solutions \cite{kingmaIntroductionVariationalAutoencoders2019}, by adding extra ROI channels to the condition input.
	
	We demonstrated the approach with a tandem network on the inverse design of dielectric core-shell nano-spheres, with the goal to separately tailor electric and magnetic dipole and/or quadrupole scattering in user-defined spectral windows.
	We showed that the network is capable to process very different design targets, ranging from the design of multipole relative spectral positions, over multipole scattering ratios and the spectral slopes to the tailoring of resonance widths. 
	
	The region-of-interest based inverse design network offers a great flexibility in inverse design and we foresee that our technique will be extremely useful in applications like metasurface design.
	Due to its broad applicability, we believe that our approach will find large interest for various further applications and also in combination with different types of network architectures.

	\section*{Acknowledgments}
	
	We thank Nathalie Destouches, Otto Muskens, Arnaud Arbouet and Christian Girard for fruitful discussions.
	A.E.-R. thanks the Institute of Quantum Technology in Occitanie IQO and the Université Paul Sabatier Toulouse for an UPS excellence PhD grant.
	This work was supported by the Toulouse HPC CALMIP (grant p20010).

	\section*{Conflicts of interests}
	
	The authors declare no competing financial interests.

	%
	

\end{document}